# Survey of excited state neutron spectroscopic factors for Z=8-28 nuclei

M.B. Tsang(曾敏兒)[1,2,3], Jenny Lee(李曉菁)[1,2], S.C. Su(蘇士俊)[4], J.Y. Dai(戴家琰)[4], M. Horoi[5], H. Liu(刘航)[1], W.G. Lynch(連致標)[1,2,3], S. Warren[3]

[1]*National Superconducting Cyclotron Laboratory, Michigan State University, East Lansing, Michigan 48824, USA*

[2]*Department of Physics and Astronomy, Michigan State University, East Lansing, Michigan 48824, USA*

[3]*Joint Institute of Nuclear Astrophysics, Michigan State University, East Lansing, Michigan 48824, USA*

[4] *Physics Department, Chinese University of Hong Kong, Shatin, Hong Kong, China*

[5]*Department of Physics, Central Michigan University, Mount Pleasant, Michigan 48859, USA*

Abstract

We have extracted 565 neutron spectroscopic factors of sd and fp shell nuclei by systematically analyzing more than 2000 measured (d,p) angular distributions. We are able to compare 125 of the extracted spectroscopic factors to values predicted by large-basis shell-model calculations and evaluate the accuracies of spectroscopic factors predicted by different shell-model interactions in these regions. We find that the spectroscopic factors predicted for most excited states of sd-shell nuclei using the latest USDB or USDA interactions agree with the experimental values. For fp shell nuclei, the inability of the current models to account for the core excitation and fragmentation of the states leads to considerable discrepancies. In particular, the agreement between data and shell-model predictions for Ni isotopes is not better than a factor of two using either the GXPF1A or the XT interaction.

Nuclear structure reflects the interplay of single-particle and collective dynamics. The occupancies and energies of the relevant single particle orbits in a given nucleus are especially important because they are essential input to large-basis shell-model (LB-SM) calculations that provide the most detailed descriptions of single-particle and collective features of medium mass sd and fp shell nuclei [1,2]. The ordering and occupancies of orbits in many interesting and important beta unstable nuclei are not known, but must be determined by measurements.

Spectroscopic factors (SF) quantify the nature and occupancy of the single-particle orbits in a nucleus, and are required to determine the orbital energies [3-5]. They provide necessary checks of the Hilbert spaces used in current nuclear structure calculations that aim to elucidate the evolution of shell structure from stable isotopes towards the drip lines. Single-nucleon spectroscopic factors also correspond to the nuclear matrix elements that describe the capture or emission of single-nucleons in stellar burning processes [6]. In the rapid neutron or proton capture processes in explosive environments, these captures often involve short-lived nuclear states with small spectroscopic factors, for which measurements can be difficult or even impossible. In such cases, shell-model calculations often provide the principal means to estimate these spectroscopic factors and the associated astrophysical rates.

To address the accuracy of such predictions, it is important to assess the accuracy of calculated spectroscopic factors using different Hilbert spaces [7]. Such an assessment may be obtained by comparing the calculated values to those extracted using well-calibrated experimental probes. In this paper, we test the predictions of LB-SM calculations [8,9] by studying SF values extracted for neutron "particle" states populated by (d,p) reactions.

A recent comparison of measured ground state neutron spectroscopic factors for Li to Cr isotopes to those calculated using large-basis shell-model calculations found that both agreed to within 20% [10-12]. This result involved systematic comparisons of the angular distribution data to Adiabatic Distorted Wave Approximation (ADWA) model calculations [13] of the transfer cross-sections [10-12]. The ADWA model addresses deuteron break-up, which can be significant for deuteron center of mass energies above 10 MeV per nucleon, by an appropriate choice of the deuteron elastic scattering potential.

Accordingly, both deuteron and proton elastic scattering optical potentials can be directly obtained from nucleon global optical potentials. In ref. [10-12] and in the present work, we used the global potentials described in ref. [14]. For simplicity, the transferred neutron was bound in the nucleus in a potential with a Woods-Saxon shape with fixed radius parameter of 1.25 fm and a diffuseness parameter of 0.65 fm. The depth of this potential is normalized to the experimental binding energy.

In general, the shell model describes the properties of ground state nuclei very well [1,2]. In astrophysics calculations, states involving resonances near the nucleon thresholds are also relevant [6] but the success of the shell model is less certain in describing such excited states [7]. To examine how well the shell model predicts the spectroscopic factors of excited states, we adopt the analysis procedure described in refs. [10-12] to extract the neutron spectroscopic factors of the excited states of the following sd shell nuclei: $^{17}$O, $^{18}$O, $^{21}$Ne, $^{24}$Na, $^{25}$Mg, $^{26}$Mg, $^{27}$Mg, $^{29}$Si, $^{30}$Si, $^{31}$Si, $^{33}$S, $^{35}$S. We also extend our analysis to $^{41,43,45,47,49}$Ca, $^{47,49,51}$Ti and $^{51,53,55}$Cr isotopes as well as $^{57,59,61,62,63,65}$Ni isotopes with neutrons in the pf shell. Comparison of experimental and theoretical spectroscopic factors provides an independent method to evaluate the interactions used in shell-model calculations, most of which have been obtained from fitting the binding energy and excitation energies of a range of nuclei.

In ref. [2], two new interactions, USDA and USDB, have been obtained to describe sd shell nuclei with an inert 16O core. In extracting the USDA and USDB interactions, constraints based on the binding energy and energy levels were used and the rms deviation of the predicted energy levels ranges from 130-170 keV. To further test the validity of these interactions, we compare the experimental and calculated spectroscopic factors, which were not used to determine the parameters of the USDA or USDB interactions. This comparison includes all (d,p) transfer reaction data on these nuclei for which spectroscopic factors can be calculated in the corresponding Hilbert spaces in large-basis shell-model (LB-SM). A detailed compilation of the experimentally extracted SFs will be provided in forthcoming publications [15, 16]. In this article, we show a quantitative overall evaluation of the success of LB-SM calculations in describing the SFs. There are only a handful of states where the agreement between the experimentally extracted and predicted SFs is unusually poor. These states are identified and discussed.

Figure 1 shows the comparisons of the experimental SFs to shell-model predictions. The averaged SF values obtained with USDA and USDB (x co-ordinates) are plotted vs. the experimental extracted SF values obtained in this study labeled as SF(ADWA) (y co-ordinates). The horizontal error bars indicate the range of the USDA and USDB results. For most of the cases, the two values are nearly identical. The solid diagonal line indicates perfect agreement between theoretical predictions and experimental data. Nearly all the data cluster around the solid line. For excited states SF, the measured angular distributions are often of poorer quality than those for the ground state. Even though we adopted the deduced experimental uncertainties of 20-30% from ref. [10,11] in the figure, the larger scatter of the excited state data could imply larger experimental uncertainties closer to 40%. For reference, the dashed lines in all the figures indicate 40% deviation from the solid line. There are three states (3.908 MeV ($5/2^+$) state in $^{25}$Mg, 7.692 MeV ($3/2^+$) and 8.290 MeV ($5/2^+$) states in $^{29}$Si), with very small calculated spectroscopic factors (<0.005), outside the range of the established systematics. Small calculated SFs originate from large cancellations of contributions from different components of the wave functions, which are hard to control even in the best shell-model calculation. Indeed, the calculated values using USDA and USDB interactions differ from each other by more than a factor of two and underpredict the experimental values by more than a factor 10. Clearly, these cases would be important to examine further, both experimentally and theoretically as the capability of predicting very small (<0.005) spectroscopic factors of sd shell nuclei is important for astrophysical applications.

Beyond the sd shell nuclei, regions of interest will be around the N=20, 28 and Z=20, 28 magic shell closures. The ground states of Ca isotopes are good single-particle states with doubly magic cores [10-11]. Spectroscopic factor predictions by both the Independent Particle Model and by the fp shell LB-SM are nearly the same and agree with the experimental values to within 20% [10-11]. Figure 2 compares excited states neutron spectroscopic factors for $^{41,43,45,47,49}$Ca, $^{47,49,51}$Ti and $^{51,53,55}$Cr isotopes. Most of the values lie within the experimental uncertainties of 40% (indicated by the deviations of the dashed lines from the solid line). On the other hand, the calculated and measured SFs near the boundaries of the fp shell-model space can disagree by factors of hundreds. The largest discrepancies when using the modern GXPF1A interaction [17,18] occur for the 2.462 MeV ($p_{3/2}$) and 6.870 MeV ($f_{5/2}$) states in $^{41}$Ca, the 2.944 MeV ($p_{3/2}$) state in $^{43}$Ca and the 4.312 MeV ($p_{1/2}$) and 4.468 MeV ($p_{1/2}$) states in $^{45}$Ca, most of which have theoretical predicted spectroscopic factors near unity. Due to proximity of these nuclei to the sd shell, their excited state wave functions have strong contributions from particle-hole excitations that lie outside the fp model space [19]. It is rather difficult at the present time to include hole excitations of the sd shell core due to the huge model space that would require reliable effective interactions for the larger sd⊗fp shell Hilbert space. In contrast, the excited states of mid-shell nuclei, such as Ti and Cr, do not have this problem and are consequently better described by the shell model.

In the case of Ni isotopes, the proton number Z=28 is a magic core. shell-model calculations for Ni isotopes assuming $^{56}$Ni as an inert core were performed in 1960's [20, 21]. Modern interactions in the fp-shell include GXPF1A [17,18] and KB3G [22]. A new T=1 effective interaction (XT) for the $f_{5/2}$, $p_{3/2}$, $p_{1/2}$, $g_{9/2}$ model space has been obtained by fitting the experimental data of Ni isotopes from A=57 to A=78 and N=50 isotones for $^{89}$Cu to $^{100}$Sn [23]. Calculations with the XT interaction, which assume a closed $^{56}$Ni core, are very fast compared to calculations with the GXPF1A interaction, which require the complete basis in the fp model space with a closed $^{40}$Ca core. Fig. 3 shows comparisons of predictions using the GXPF1A interaction (left panel) and the XT interaction (right panel) to experimental values. The current analysis approach yields spectroscopic factors that cluster around the large-basis shell-model predictions. Due to difficulties in identifying states at higher excitation energy, only SF values for a few states less than 3 MeV are obtained from calculations using the GXPF1A interaction. More states from calculations using the XT interaction can be compared to data as shown in the right panel. The data and the predictions agree to within a factor of two. The scatter in the calculated values is large compared to experimental uncertainties, which are estimated to be around 40% as indicated by the dashed lines above and below the solid line. For small calculated SF values of less than 0.01, two of the data points (the 4.709 MeV, 9/2$^+$ state in $^{59}$Ni and the 2.124 MeV, 1/2$^-$ state in $^{61}$Ni) deviate from the systematics in the right panel. Unfortunately, there are not enough statistics to draw a firm conclusion on the reliability of small calculated SF values from calculations with the XT interaction in the Ni isotopes.

We have used the established systematics between the experimental and theoretical spectroscopic factors to assign the spins of three selected states in $^{27}$Mg that have no definitive spin assignments from measurements. The spins for these states listed in Table 1 can either be $3/2^+$ or $5/2^+$ (second column in Table 1) according to NUDAT [24]. Since the measured angular distributions are sensitive to the angular momentum $l$ but not very sensitive to the spin J, the extracted SF values (third column) are similar for different values of J. However, the shell-model spectroscopic factors (sixth column) for the $3/2^+$ and $5/2^+$ states within 100 keV of the 5.627 MeV state differ by more than a factor of 25. The systematics of Figure 1 indicates that the spins of the 5.627 MeV, 3.491 MeV and 4.150 MeV states are consistent with a J=$3/2^+$, J=$3/2^+$ and J=$5/2^+$, respectively.

In summary, spectroscopic factors provide an independent test for the effective interactions used in shell models. Excited-state neutron spectroscopic factors have been extracted from a range of isotopes from Z=8 to 28. The extracted values provide information on single-particle levels. For states in sd shell nuclei, spectroscopic factors calculated using USDA and USDB interactions reproduce the experimental values, within the experimental uncertainties, when the calculated values exceed 0.005. Outside the $^{40}$Ca core, the excited state spectroscopic factors of Ca, Ti and Cr isotopes are less accurate, especially for light Ca isotopes near the closed sd shell where a large sd-fp model space may be needed. In these cases, the shell model predicts purer single-particle states with larger spectroscopic factors than the experimentally observed. Away from closed shells, fragmentation of states is better predicted by the fp shell model. For the excited states in Ni isotopes, the agreement is poor and the measured spectroscopic factors of the excited states cannot distinguish whether the GXPF1A interaction in the full fp model space or the XT interaction in the $f_{5/2}p_{3/2}p_{1/2}g_{9/2}$ model space is better.

Our results indicate that the calculated spectroscopic factors correlate stronger with those extracted from experimental data when better effective interactions are used or/and when larger valence spaces can be accommodated in the calculations. Both directions require more efficient shell-model codes/algorithms and larger computational facilities, which will become available in the forthcoming years. The good agreement for small spectroscopic factors in the sd shell nuclei suggests that experiments can be reliably performed to extract SF's values down to 0.005. The ability to measure and calculate small SF values is important for many states of astrophysical interest.

**Acknowledgement**

The authors would like to thank Professor B.A. Brown, Professor W. Richter and Professor J. Tostevin for fruitful discussions and the High Performance Computer Center at MSU for making it possible to do the LB-SM calculations in Cr and Ti isotopes. This work is supported by the National Science Foundation under grants PHY-0606007, PHY 0216783, and PHY-0758099 (MH). Horoi acknowledges the NSF MRI grant PHY-0619407, which made possible the full fp model space shell-model calculations with the GXPF1A interaction. Su (2006) and Dai (2007) acknowledge the support of the Summer Undergraduate Research Experience (SURE) program sponsored by the Chinese University of Hong Kong.

Table I: Spin assignments of three excited states, 3.491, 4.150 and 5.627 MeV of $^{27}$Mg. The NUDAT [24] values listed in column 2 are not confirmed by experiments. The shell-model information is listed in Column 4, 5 and 6. Our recommended spin values are listed in the last column.

| E* (MeV) | J (NUDAT) | SF(ADWA) | E(LB-SM) | J(LB-SM) | SF(LB-SM) | SF(ADWA)/SF(LB-SM) | J (this work) |
|---|---|---|---|---|---|---|---|
| **3.49** | (3/2+) | 0.049 ± 0.015 | **3.562** | 3/2+ | 0.097 ± 0.012 | 0.51±0.17 | **3/2+** |
|  | (5/2+) | 0.032 ± 0.010 |  |  |  |  |  |
|  |  |  |  |  |  |  |  |
| **4.15** | (3/2+) | 0.038 ± 0.011 |  |  |  |  |  |
|  | (5/2+) | 0.025 ± 0.007 | **4.097** | 5/2+ | 0.029 ± 0.002 | 0.86±0.25 | **5/2+** |
|  |  |  |  |  |  |  |  |
| **5.627** | (3/2+) | 0.129 ± 0.039 | **5.561** | 3/2+ | 0.144 ± 0.003 | 0.89±0.27 | **3/2+** |
|  | (5/2+) | 0.085 ± 0.026 | **5.690** | 5/2+ | 0.0054 ± 0.0004 | 16±4.8 |  |

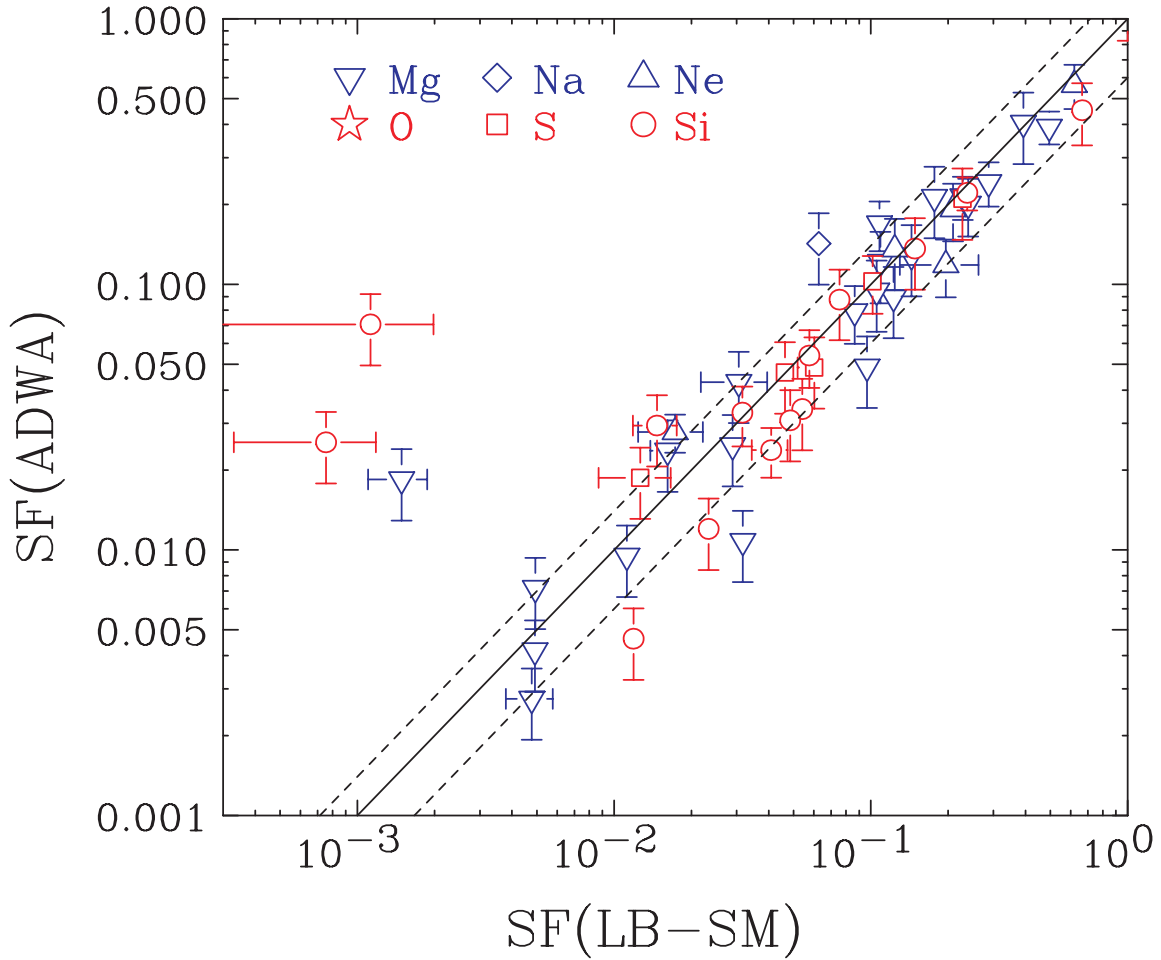

Figure 1 (Color online): Comparison of experimental excited states spectroscopic factors, SF(ADWA), to predictions from large-basis shell-model calculations, SF(LBSM), using the USDA and USDB interactions. The ends of the horizontal error bars indicate the range of values predicted by USDA and USDB interactions. Symbols indicate the averaged values. The solid line represents perfect agreement between data and theory. The dashed lines correspond to ±40% deviations (expected experimental uncertainties) from the solid line.

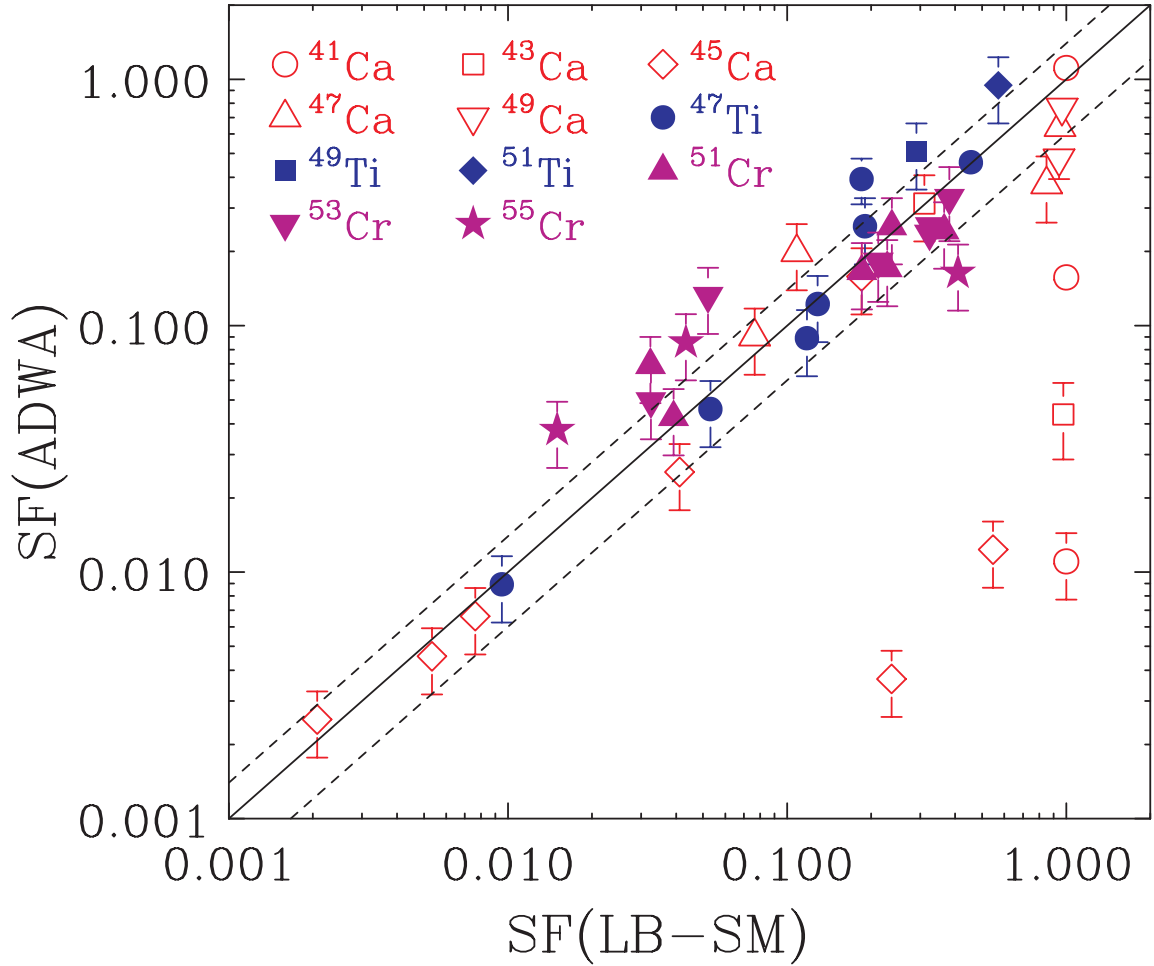

Figure 2 (Color online): Comparison of experimental spectroscopic factors, SF(ADWA), to predictions from large-basis shell-model for the Ca, Ti and Cr isotopes, SF(LB-SM). Complete basis with the interaction GXPF1A is used in the theoretical calculations. The solid line indicates perfect agreement between data and predictions. Dashed lines represent ±40% deviations from the solid line.

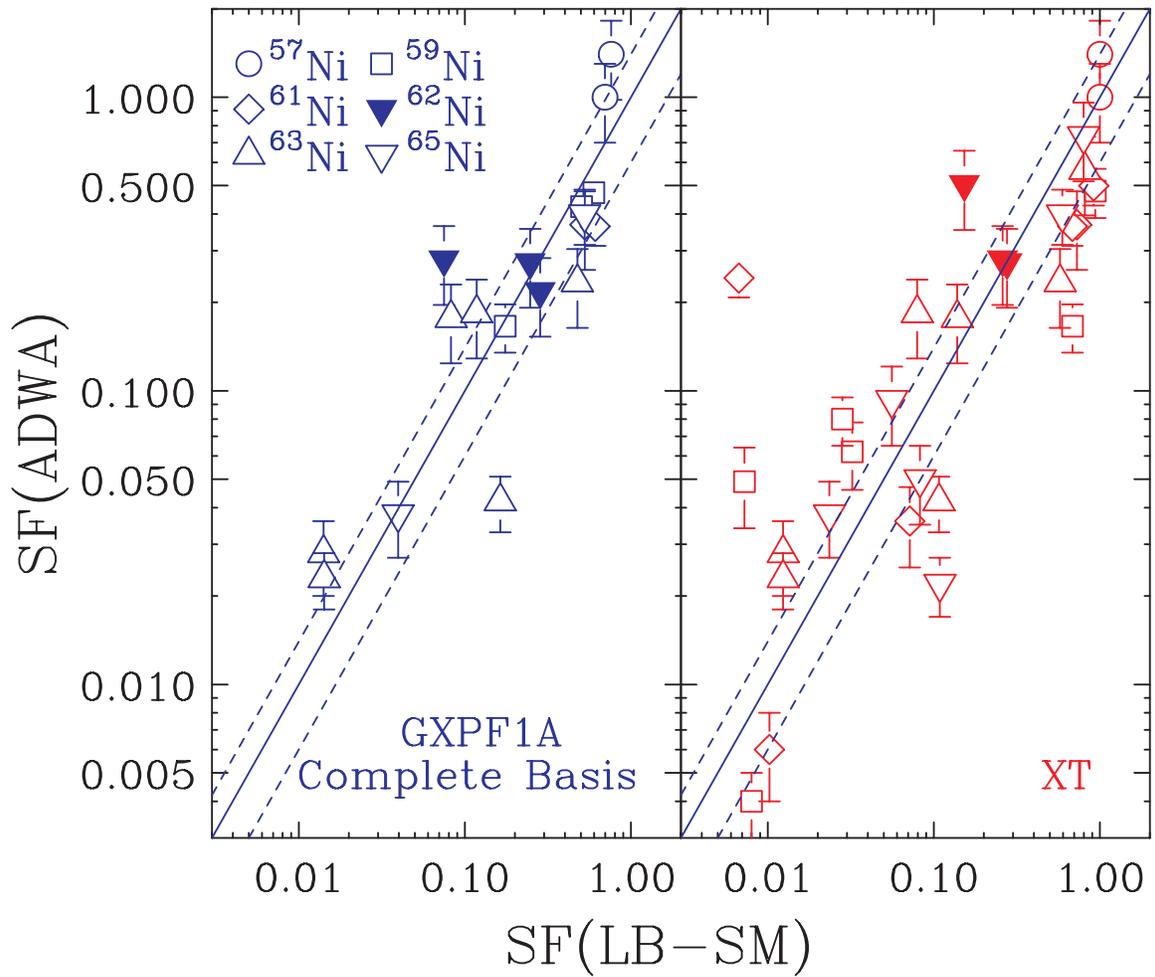

Fig. 3 (Color online): Comparison of the experimental SF values, SF(ADWA) and the shell-model calculations with the GXPF1A interaction in the pf model space (left panel) and the XT interaction in gfp model space (right panel). The solid line indicates perfect agreement between data and predictions. Dashed lines represent ±40% deviations from the solid line.